\documentclass[10pt,conference,a4paper]{IEEEtran}
\IEEEoverridecommandlockouts
\usepackage{cite}
\usepackage{amsmath,amssymb,amsfonts}
\usepackage{algorithmic}
\usepackage{graphicx}
\usepackage{textcomp}
\usepackage{xcolor}
\usepackage{pgfplots}
\usepackage{float}
\usepackage{bm}
\usepackage{hyperref}
\usepackage[utf8]{inputenc}
\DeclareUnicodeCharacter{2212}{−}
\usepgfplotslibrary{groupplots,dateplot}
\usetikzlibrary{patterns,shapes,arrows,arrows.meta}
\pgfplotsset{compat=newest}
\usetikzlibrary{positioning}
\graphicspath{{figures/}}
\def\BibTeX{{\rm B\kern-.05em{\sc i\kern-.025em b}\kern-.08em
    T\kern-.1667em\lower.7ex\hbox{E}\kern-.125emX}}

\DeclareMathOperator*{\argmin}{\arg\!\min}

\newcommand{\SmoothingMatrixInvFactor}{\alpha}
\newcommand{\BSRecon}{B_\text{r}}
\newcommand{\BSCSR}{B_\text{csr}}
\newcommand{\ImageHeight}{H}
\newcommand{\NumMSChannels}{M}
\newcommand{\NumHSChannels}{N}
\newcommand{\NumStackedBlocks}{T}
\newcommand{\ImageWidth}{W}

\newcommand{\IntensityLevel}{l}
\newcommand{\CameraSpectrum}{m}
\newcommand{\LensSpectrum}{o}

\newcommand{\LightSourceSpectrum}{q}
\newcommand{\ReflectanceSpectrum}{r}
\newcommand{\SpectrumFunction}{s}
\newcommand{\Wavelength}{\lambda}
\newcommand{\MaxBlocks}{\mu_{\mathcal{C}}}
\newcommand{\ThresholdBlocks}{\tau}
\newcommand{\ThresholdBlocksConstant}{\tau_{\mathcal{C}}}
\newcommand{\ImageCoords}{\Omega}
\newcommand{\Indicator}{\xi}
\newcommand{\MSChannels}{\bm{c}}
\newcommand{\MSChannelsBig}{\bm{\hat{c}}}
\newcommand{\SingleFilter}{\bm{f}}
\newcommand{\Noise}{\bm{n}}
\newcommand{\NoiseBig}{\bm{\hat{n}}}
\newcommand{\HSChannels}{\bm{s}}
\newcommand{\HSChannelsBig}{\bm{\hat{s}}}
\newcommand{\HSChannelsEst}{\bm{\check{s}}}
\newcommand{\NoiseVar}{\bm{\sigma}^2}
\newcommand{\CoordOne}{\bm{x}}
\newcommand{\CoordTwo}{\bm{y}}
\newcommand{\MSImage}{\bm{C}}
\newcommand{\MSImageBlockVec}{\bm{C}_{\text{b}}}
\newcommand{\MSImageNoiseless}{\bm{\check{C}}}
\newcommand{\DifferenceMatrix}{\bm{D}}
\newcommand{\FilterMatrix}{\bm{F}}
\newcommand{\BigFilterMatrix}{\bm{\hat{F}}}
\newcommand{\CorrelationMS}{\bm{K}_{\text{c}}}
\newcommand{\CorrelationHS}{\bm{K}_{\text{s}}}
\newcommand{\CorrelationHSMS}{\bm{K}_{\text{sc}}}

\newcommand{\CorrelationSpatioSpectralCSR}{\bm{K}_{\text{CSR}}}
\newcommand{\Identity}{\bm{I}}
\newcommand{\SmoothingMatrix}{\bm{M}}
\newcommand{\MSNoiseVarMatrix}{\bm{N}}
\newcommand{\MSNoiseVarMatrixBig}{\bm{\hat{N}}}

\newcommand{\HSImage}{\bm{S}}
\newcommand{\HSImageEst}{\bm{\hat{S}}}

\newcommand{\BlockIndices}{\bm{Z}}

\newcommand{\MSCube}{\mathcal{C}}
\newcommand{\HSCube}{\mathcal{S}}
\newcommand{\HSCubeExtended}{\mathcal{\hat{S}}}

\newcommand{\fig}{Fig.\,}

\newcommand{\overbar}[1]{\mkern 1.5mu\overline{\mkern-1.5mu#1\mkern-1.5mu}\mkern 1.5mu}

\makeatletter
\def\ps@IEEEtitlepagestyle{%
	\def\@oddfoot{\mycopyrightnotice}%
	\def\@oddhead{\hbox{}\@IEEEheaderstyle\leftmark\hfil\thepage}\relax
	\def\@evenhead{\@IEEEheaderstyle\thepage\hfil\leftmark\hbox{}}\relax
	\def\@evenfoot{}%
}
\def\mycopyrightnotice{%
	\begin{minipage}{\textwidth}
		\scriptsize
		\copyright 2021 IEEE.  Personal use of this material is permitted. Permission from
		IEEE must be obtained for all other uses, in any current or future
		media, including reprinting/republishing this material for advertising
		or promotional purposes, creating new collective works, for resale or
		redistribution to servers or lists, or reuse of any copyrighted
		component of this work in other works. DOI: \url{https://doi.org/10.1109/MMSP53017.2021.9733655}
	\end{minipage}
}
\makeatother

\IEEEoverridecommandlockouts
\IEEEpubid{\makebox[\columnwidth]{978-1-6654-3288-7/21/\$31.00~\copyright2021 IEEE \hfill} \hspace{\columnsep}\makebox[\columnwidth]{ }}

\begin{document}

\title{Hyperspectral Image Reconstruction from Multispectral Images Using Non-Local Filtering}

\author{\IEEEauthorblockN{Frank Sippel, Jürgen Seiler, and André Kaup}
\IEEEauthorblockA{\textit{Multimedia Communications and Signal Processing} \\
\textit{Friedrich-Alexander University Erlangen-Nürnberg (FAU)}\\
Cauerstr. 7, 91058 Erlangen, Germany \\
\{frank.sippel, juergen.seiler, andre.kaup\} @fau.de}
}

\maketitle
\IEEEpubidadjcol

\begin{abstract}
Using light spectra is an essential element in many applications, for example, in material classification. Often this information is acquired by using a hyperspectral camera. Unfortunately, these cameras have some major disadvantages like not being able to record videos. Therefore, multispectral cameras with wide-band filters are used, which are much cheaper and are often able to capture videos. However, using multispectral cameras requires an additional reconstruction step to yield spectral information. Usually, this reconstruction step has to be done in the presence of imaging noise, which degrades the reconstructed spectra severely. Typically, same or similar pixels are found across the image with the advantage of having independent noise. In contrast to state-of-the-art spectral reconstruction methods which only exploit neighboring pixels by block-based processing, this paper introduces non-local filtering in spectral reconstruction. First, a block-matching procedure finds similar non-local multispectral blocks. Thereafter, the hyperspectral pixels are reconstructed by filtering the matched multispectral pixels collaboratively using a reconstruction Wiener filter. The proposed novel procedure even works under very strong noise. The method is able to lower the spectral angle up to 18\% and increase the peak signal-to-noise-ratio up to 1.1dB in noisy scenarios compared to state-of-the-art methods. Moreover, the visual results are much more appealing.
\end{abstract}


\section{Introduction}
\label{sec:intro}

Light spectra are used in many different classification tasks, for example, classifying different types of plastic \cite{moroni-pet-2015}, determining the degree of burn \cite{sowa-classification-2006}, or detecting counterfeits in a pharmaceutical context \cite{degardin-near-2016}. Hyperspectral cameras are able to capture images at many different wavelengths, typically from 25 up to several thousand bands. Since these bands are usually uniformly distributed in the wavelength area of interest, the result is a sampled spectrum for each pixel.

\begin{figure}[tb]
	\begin{center}
		\includegraphics[scale=0.8]{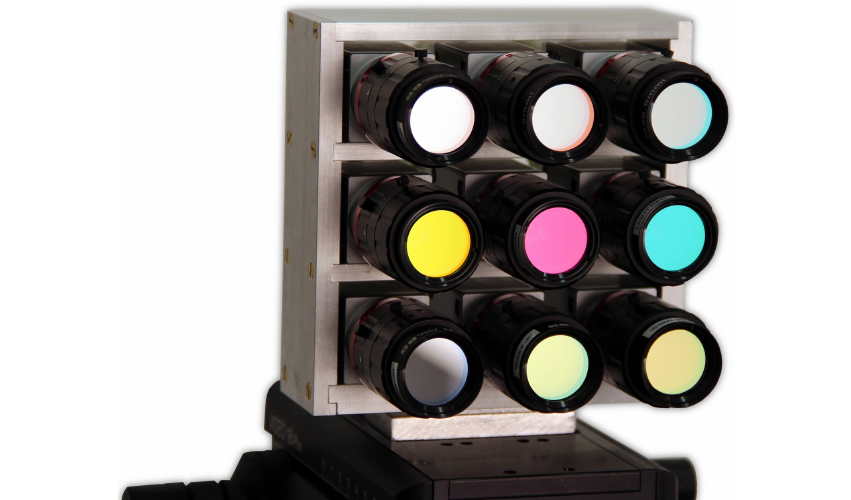}
	\end{center}
	\vspace*{-0.2cm}
	\caption{A multispectral camera \cite{genser_camera_2020}.}
	\label{fig:ms_camera}
\end{figure}

Unfortunately, \textit{hyperspectral} cameras are usually expensive and either suffer from a low spatial resolution or are not able to capture videos as it is the case for push broom cameras. This can be alleviated by deploying \textit{multispectral} cameras. Multispectral cameras typically record a scene using around six to 16 wide-band filters. Such a camera with nine channels is shown in \fig\ref{fig:ms_camera} \cite{genser_camera_2020}. Therefore, the basic spectral resolution from a multispectral camera is lower than from a hyperspectral camera. However, the full spectrum can be estimated based on the multispectral channels and prior information about the spectra in general \cite{pratt-spectral-1976}. Thus, by exploiting the overlapping parts of the transfer functions of the spectral filters, the underlying spectrum can be reconstructed. For this, the spectral transfer functions of the filters must be known.

A challenge for the spectral reconstruction is noise. Typically, the main source of noise in imaging pipeline results from miscounting photons, the so-called shot noise \cite{schoberl-photometric-2012}. The less photons available to count, the higher the relative number of photons that are missed. There are several sources which limit the amount of photons that can be counted by the sensor. In a dark environment the available photons are limited by definition. Moreover, the bandpass filters block most of photons that would reach the sensor. All of this can basically be compensated by increasing the exposure time. However, when recording a video or imaging a dynamic scene, the exposure time has a hard physical limit. Hence, in multispectral imaging and especially multispectral video recording, noise is a severe problem that influences the spectral reconstruction strongly negatively.

There are different approaches to solve this inverse problem in the presence of noise, for example a Wiener filter \cite{pratt-spectral-1976}. However, these often do only consider a small amount of noise. Other algorithms also consider stronger noise and embed spatial correlation into the reconstruction. For example the spatio-spectral Wiener filter (SSW) \cite{murakami-color-2008} tries to lower the influence of noise by modeling the neighborhood in vertical and horizontal direction as a first-order Markov process. Another approach is to use guided filtering to improve the reconstruction result, which is called structure-preserving reflectance estimation (SPRE) \cite{sippel-structure-preserving-2020}. SPRE and SSW perform well for a medium amount of noise, however in environments with strong noise they also fail to deliver good results in terms of measurable error, as well as from a visual point of view. Both methods only use the local neighborhood to encounter noise, although it is known that non-local similarities can improve denoising tasks significantly \cite{dabov-image-2007}. Thus, a novel spectral reconstruction method for noisy environments is proposed that reconstructs globally similar pixels collaboratively to encounter noise.

The problem statement of this paper is described in Section \ref{sec:problem_statement}, while the novel algorithm is described in Section \ref{sec:csr}. The methods are evaluated in Section \ref{sec:evaluation}. Finally, the results are summarized in Section \ref{sec:conclusion}.

\section{Problem Statement}
\label{sec:problem_statement}
During the imaging process, the $i$-th of $\NumMSChannels$ multispectral channels $\MSChannels_i$ can be modeled by counting photons, thus integrating over the considered spectral range from $\Wavelength_\text{min}$ to $\Wavelength_\text{max}$ \cite{cortes-multipectral-2003}
\begin{equation}
	\label{eq:basic_eq}
	\MSChannels_i = \int_{\Wavelength_\text{min}}^{\Wavelength_\text{max}} \LightSourceSpectrum(\Wavelength)\ReflectanceSpectrum(\Wavelength)\SingleFilter_i(\Wavelength)\LensSpectrum(\Wavelength)\CameraSpectrum(\Wavelength) \ \text{d}\Wavelength,
\end{equation}
where $\LightSourceSpectrum(\Wavelength)$ is the spectrum of the light source, $\ReflectanceSpectrum(\Wavelength)$ is the spectrum that is reflected by the imaged object, $\SingleFilter_i(\Wavelength)$ is the transfer function of the $i$-th filter of the multispectral camera, $\LensSpectrum(\Wavelength)$ is the absorption spectrum of the lens and $\CameraSpectrum(\Wavelength)$ is the spectral sensitivity of the camera. $\LightSourceSpectrum(\Wavelength)$, $\ReflectanceSpectrum(\Wavelength)$, $\LensSpectrum(\Wavelength)$ and $\CameraSpectrum(\Wavelength)$ are unknown. Thus, effectively only the spectrum ${\SpectrumFunction(\Wavelength) = \LightSourceSpectrum(\Wavelength)\ReflectanceSpectrum\left(\Wavelength\right)\LensSpectrum(\Wavelength)\CameraSpectrum(\Wavelength)}$ can be reconstructed from the recorded multispectral channels $\MSChannels$.

Since inverting an integration is cumbersome, the spectrum $\SpectrumFunction(\Wavelength)$ and the filters $\SingleFilter_i(\Wavelength)$ are sampled with $\NumHSChannels$ sampling points, which results in ${\HSChannels = \left[\SpectrumFunction(\Wavelength_\text{1}), \SpectrumFunction(\Wavelength_\text{2}), \cdots, \SpectrumFunction(\Wavelength_\NumHSChannels)\right] \in \mathcal{R}^\NumHSChannels}$ and ${\FilterMatrix_i=\left[\SingleFilter_i(\Wavelength_\text{1}), \SingleFilter_i(\Wavelength_\text{2}), \cdots, \SingleFilter_i(\Wavelength_\NumHSChannels)\right]}$, respectively, where $\FilterMatrix_i$ corresponds to the $i$-th row of the filter matrix ${\FilterMatrix \in \mathcal{R}^{\NumMSChannels \times \NumHSChannels}}$. Using these sampled versions, one can replace the integration in \eqref{eq:basic_eq} by a sum, which results in the underdetermined linear system 
\begin{equation}
\MSChannels = \FilterMatrix \HSChannels
\end{equation}
for the noiseless case.

Unfortunately, each image acquisition is affected by imaging noise. For example, the noise results from a limited exposure time when capturing videos. There are several types of noise in the imaging pipeline. According to \cite{schoberl-photometric-2012}, under low light conditions shot noise is the most important noise source. It is Poisson distributed, which is inconvenient to model since it often ends in non-closed-form solutions. For the derivation of a closed-form solution, this noise source is therefore often approximated by additive white Gaussian noise (AWGN), which results in
\begin{equation}
	\MSChannels = \FilterMatrix \HSChannels + \Noise,
\end{equation}
where $\Noise$ is an AWGN source which is independently applied to every multispectral channel. It is noteworthy that during testing Poisson noise is used to simulate a realistic environment.

The goal of this paper is to reduce the influence of this noise on the reconstruction. Though the noise itself is unknown, noise statistics are helpful determining the strength of the noise. Since AWGN is assumed, the strength of noise is measured by the noise variance, which can be estimated by \cite{immerkaer-fast-1996}. Since the noise of individual multispectral channels is independent of each other and of zero mean, the noise covariance matrix $\MSNoiseVarMatrix = \mathcal{E}\{ \Noise \Noise^{\text{T}} \}$ is just a diagonal matrix with the corresponding noise variance on the diagonal.

To encounter noise, spatial correlation is exploited by reconstructing several pixels at once. Therefore, the single-pixel equations above need to be extended to whole multispectral images $\MSImage(\CoordOne)$ and hyperspectral images $\HSImage(\CoordOne)$, where $\CoordOne$ is the two-dimensional image coordinate. Often cubes of multispectral images are processed. Therefore, $\MSImageBlockVec^{\CoordOne}$ is the multispectral cube with center pixel $\CoordOne$. Furthermore, the filter matrix needs to be extended $\BigFilterMatrix = \Identity \otimes \FilterMatrix$, where $\Identity$ is the identity matrix and $\otimes$ is the Kronecker product. The size of $\Identity$ depends on the amount of pixels to reconstruct at once. Similarly, the noise vector has to be extended to match the dimensions. Since the noise is unknown, only the size of the vector is extended. Thus, the size of the extended noise vector $\NoiseBig$ is the multiplication of the multispectral channels with the amount of pixels that are reconstructed at once. Hence, \eqref{eq:basic_eq} extends to
\begin{equation}
	\label{eq:basic_eq_extended}
	\MSChannelsBig = \BigFilterMatrix \HSChannelsBig + \NoiseBig,
\end{equation}
where $\MSChannelsBig$ and $\HSChannelsBig$ are the extended multispectral channels and hyperspectral channels, respectively. These contain the data of several pixels. Moreover, the noise covariance matrix $\MSNoiseVarMatrix$ also needs to be extended, which results in $\MSNoiseVarMatrixBig = \Identity \otimes \MSNoiseVarMatrix$. 

\section{Proposed Method}
\label{sec:csr}

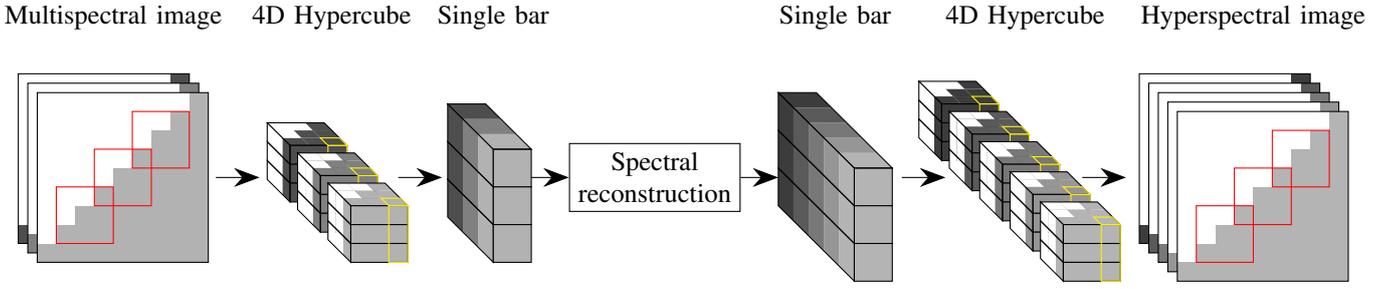
\begin{figure*}[tb]
	\begin{center}
		\begin{tikzpicture}[scale=0.25]
			\begin{scope}[shift={(0, -1)}]
				\def\grayarray{{178, 128, 78}}
				\foreach \i in {2,...,0}
				{
					\foreach \x in {0,...,8}
					{
						\foreach \y in {0,...,8}
						{
							\ifnum \y>\x
							\definecolor{fillcolor}{RGB}{255, 255, 255}
							\else
							\pgfmathparse{\grayarray[\i]};
							\definecolor{fillcolor}{RGB}{\pgfmathresult, \pgfmathresult, \pgfmathresult}
							\fi
							
							\fill[fillcolor] (1 - \i * 0.5 + \x,\i * 0.5 + \y) rectangle (1 - \i * 0.5 + \x + 1, \i * 0.5 + \y + 1);
						}
					}
					\draw[draw=black] (1 - \i * 0.5,\i * 0.5) rectangle (1 - \i * 0.5 + 9,\i * 0.5 + 9);
				}
				
				\draw[draw=red] (2,1) rectangle (5,4);
				\draw[draw=red] (4,3) rectangle (7,6);
				\draw[draw=red] (6,5) rectangle (9,8);
				
				\draw[-{Stealth[scale=2]}] (10.4, 4.5) -- (12.7, 4.5);
				
				\begin{scope}[shift={(17.5, 0)}, x={(1, 0)}, y={(0, 1)}, z={(-0.4, 0.4)}]
					\def\height{3}
					\foreach \i in {2,...,0}
					{
						\begin{scope}[shift={(0, 0, \i * 4)}]
							\foreach \x in {0,...,2}
							{
								\foreach \z in {0,...,2}
								{
									\ifnum \z>\x
									\definecolor{fillcolor}{RGB}{255, 255, 255}
									\else
									\pgfmathparse{\grayarray[\i]};
									\definecolor{fillcolor}{RGB}{\pgfmathresult, \pgfmathresult, \pgfmathresult}
									\fi
									\fill[fill=fillcolor] (\x, \height, 0 + \z) -- (1 + \x, \height, 0 + \z) -- (1 + \x, \height, 1 + \z) -- (\x, \height, 1 + \z) -- cycle;
								}
							}
							\draw[] (0, \height, 0) -- (3, \height, 0) -- (3, \height, 1 + 2) -- (0, \height, 1 + 2) -- cycle;
							
							\pgfmathparse{\grayarray[\i]};
							\definecolor{fillcolor}{RGB}{\pgfmathresult, \pgfmathresult, \pgfmathresult}
							\foreach \j in {0,...,2}
							{
								\ifnum \j > 0
								\definecolor{fillcolor}{RGB}{255, 255, 255}
								\else
								\pgfmathparse{\grayarray[\i]};
								\definecolor{fillcolor}{RGB}{\pgfmathresult, \pgfmathresult, \pgfmathresult}
								\fi
								\fill[fill=fillcolor] (0, 0, \j) -- (0, 0, \j + 1) -- (0, \height, \j + 1) -- (0, \height, \j) -- cycle;
							}
							\foreach \j in {0,...,2}
							{
								\draw[fill=fillcolor] (0, \height - \j - 1, 0) -- (3, \height - \j - 1, 0) -- (3, \height - \j, 0) -- (0, \height - \j, 0) -- cycle;
								
								\draw[] (0, \height - \j - 1, 0) -- (0, \height - \j - 1, 3) -- (0, \height - \j, 3) -- (0, \height - \j, 0) -- cycle;
							}
							
							\draw[draw=yellow] (2, 0, 0) -- (2, \height, 0) -- (3, \height, 0) -- (3, 0, 0) -- cycle;
							\draw[draw=yellow] (2, \height, 0) -- (2, \height, 1) -- (3, \height, 1) -- (3, \height, 0) -- cycle;
						\end{scope}
					}
				\end{scope}
				
				\draw[-{Stealth[scale=2]}] (20, 4.5) -- (22.3, 4.5);
				
				\begin{scope}[shift={(25, 0)}, x={(1, 0)}, y={(0, 1)}, z={(-0.4, 0.4)}]
					\def\height{6}
					\foreach \i in {0,...,2}
					{
						\foreach \j in {0,...,2}
						{
							\pgfmathparse{\grayarray[\i]};
							\definecolor{fillcolor}{RGB}{\pgfmathresult, \pgfmathresult, \pgfmathresult}
							
							\fill[fill=fillcolor] (0, \height - \j * 2 - 2, \i * 2) -- (0, \height - \j * 2 - 2, \i * 2 + 2) -- (0, \height - \j * 2, \i * 2 + 2) -- (0, \height - \j * 2, \i * 2) -- cycle;
						}
					}
					\foreach \i in {0,...,2}
					{
						\draw[] (0, \height - \i * 2 - 2, 0) -- (0, \height - \i * 2 - 2, 6) -- (0, \height - \i * 2, 6) -- (0, \height - \i * 2, 0) -- cycle;
						
						\pgfmathparse{\grayarray[\i]};
						\definecolor{fillcolor}{RGB}{\pgfmathresult, \pgfmathresult, \pgfmathresult}
						\fill[fill=fillcolor] (0, \height, \i * 2) -- (0, \height, \i * 2 + 2) -- (2, \height, \i * 2 + 2) -- ( + 2, \height, \i * 2) -- cycle;
						
						\pgfmathparse{\grayarray[0]};
						\definecolor{fillcolor}{RGB}{\pgfmathresult, \pgfmathresult, \pgfmathresult}
						\draw[fill=fillcolor] (0, \height - \i * 2, 0) -- (0, \height - \i * 2 - 2, 0) -- (2, \height - \i * 2 - 2, 0) -- (2, \height - \i * 2, 0) -- cycle;
					}
					
					\draw[] (0, \height, 0) -- (2, \height, 0) -- (2, \height, 6) -- (0, \height, 6) -- cycle;
				\end{scope}

				\node[] at (5,13) {Multispectral image};
				\node[] at (16.5,13) {4D Hypercube};
				\node[] at (25,13) {Single bar};
			\end{scope}
			\node[draw, align=center] at (33.5, 3.5) (sr) {Spectral\\reconstruction};
			\draw[-{Stealth[scale=2]}] (27, 3.5) -- (sr.west);
			
			\begin{scope}[shift={(40, -2)}]
				\def\grayarray{{180, 150, 120, 90, 60}}
				
				\begin{scope}[shift={(4, 0)}, x={(1, 0)}, y={(0, 1)}, z={(-0.4, 0.4)}]
					\def\height{6}
					\foreach \i in {0,...,4}
					{
						\foreach \j in {0,...,2}
						{
							\pgfmathparse{\grayarray[\i]};
							\definecolor{fillcolor}{RGB}{\pgfmathresult, \pgfmathresult, \pgfmathresult}
							
							\fill[fill=fillcolor] (0, \height - \j * 2 - 2, \i * 2) -- (0, \height - \j * 2 - 2, \i * 2 + 2) -- (0, \height - \j * 2, \i * 2 + 2) -- (0, \height - \j * 2, \i * 2) -- cycle;
						}
					}
					\foreach \i in {0,...,4}
					{	
						\pgfmathparse{\grayarray[\i]};
						\definecolor{fillcolor}{RGB}{\pgfmathresult, \pgfmathresult, \pgfmathresult}
						\fill[fill=fillcolor] (0, \height, \i * 2) -- (0, \height, \i * 2 + 2) -- (2, \height, \i * 2 + 2) -- (2, \height, \i * 2) -- cycle;
					}
					\foreach \i in {0,...,2}
					{
						\draw[] (0, \height - \i * 2 - 2, 0) -- (0, \height - \i * 2 - 2, 10) -- (0, \height - \i * 2, 10) -- (0, \height - \i * 2, 0) -- cycle;
						
						\pgfmathparse{\grayarray[0]};
						\definecolor{fillcolor}{RGB}{\pgfmathresult, \pgfmathresult, \pgfmathresult}
						\draw[fill=fillcolor] (0, \height - \i * 2, 0) -- (0, \height - \i * 2 - 2, 0) -- (2, \height - \i * 2 - 2, 0) -- (2, \height - \i * 2, 0) -- cycle;
					}
					
					\draw[] (0, \height, 0) -- (2, \height, 0) -- (2, \height, 10) -- (0, \height, 10) -- cycle;
					\node[] at (1, \height, 10) (hsbar) {};
				\end{scope}

				\draw[-{Stealth[scale=2]}] (6.5, 5.5) -- (8.8, 5.5);
				
				
				\begin{scope}[shift={(15, 0)}, x={(1, 0)}, y={(0, 1)}, z={(-0.4, 0.4)}]
					\def\height{3}
					\foreach \i in {4,...,0}
					{
						\begin{scope}[shift={(0, 0, \i * 4)}]
							\foreach \x in {0,...,2}
							{
								\foreach \z in {0,...,2}
								{
									\ifnum \z>\x
									\definecolor{fillcolor}{RGB}{255, 255, 255}
									\else
									\pgfmathparse{\grayarray[\i]};
									\definecolor{fillcolor}{RGB}{\pgfmathresult, \pgfmathresult, \pgfmathresult}
									\fi
									\fill[fill=fillcolor] (\x, \height, 0 + \z) -- (1 + \x, \height, 0 + \z) -- (1 + \x, \height, 1 + \z) -- (\x, \height, 1 + \z) -- cycle;
								}
							}
							\draw[] (0, \height, 0) -- (3, \height, 0) -- (3, \height, 1 + 2) -- (0, \height, 1 + 2) -- cycle;
							
							\pgfmathparse{\grayarray[\i]};
							\definecolor{fillcolor}{RGB}{\pgfmathresult, \pgfmathresult, \pgfmathresult}
							\foreach \j in {0,...,2}
							{
								\ifnum \j > 0
								\definecolor{fillcolor}{RGB}{255, 255, 255}
								\else
								\pgfmathparse{\grayarray[\i]};
								\definecolor{fillcolor}{RGB}{\pgfmathresult, \pgfmathresult, \pgfmathresult}
								\fi
								\fill[fill=fillcolor] (0, 0, \j) -- (0, 0, \j + 1) -- (0, \height, \j + 1) -- (0, \height, \j) -- cycle;
							}
							\foreach \j in {0,...,2}
							{
								\draw[fill=fillcolor] (0, \height - \j - 1, 0) -- (3, \height - \j - 1, 0) -- (3, \height - \j, 0) -- (0, \height - \j, 0) -- cycle;
								
								\draw[] (0, \height - \j - 1, 0) -- (0, \height - \j - 1, 3) -- (0, \height - \j, 3) -- (0, \height - \j, 0) -- cycle;
							}
							
							\draw[draw=yellow] (2, 0, 0) -- (2, \height, 0) -- (3, \height, 0) -- (3, 0, 0) -- cycle;
							\draw[draw=yellow] (2, \height, 0) -- (2, \height, 1) -- (3, \height, 1) -- (3, \height, 0) -- cycle;
						\end{scope}
					}
				\end{scope}
				
				\begin{scope}[shift={(20, 0)}]
					\foreach \i in {4,...,0}
					{
						\foreach \x in {0,...,8}
						{
							\foreach \y in {0,...,8}
							{
								\ifnum \y>\x
								\definecolor{fillcolor}{RGB}{255, 255, 255}
								\else
								\pgfmathparse{\grayarray[\i]};
								\definecolor{fillcolor}{RGB}{\pgfmathresult, \pgfmathresult, \pgfmathresult}
								\fi
								
								\fill[fillcolor] (1 - \i * 0.5 + \x,\i * 0.5 + \y) rectangle (1 - \i * 0.5 + \x + 1, \i * 0.5 + \y + 1);
							}
						}
						\draw[draw=black] (1 - \i * 0.5,\i * 0.5) rectangle (1 - \i * 0.5 + 9,\i * 0.5 + 9);
					}
					
					\draw[draw=red] (2,1) rectangle (5,4);
					\draw[draw=red] (4,3) rectangle (7,6);
					\draw[draw=red] (6,5) rectangle (9,8);
				\end{scope}
				
				\draw[-{Stealth[scale=2]}] (16, 5.5) -- (18.3, 5.5);
				
				\node[] at (25,14) {Hyperspectral image};
				\node[] at (13,14) {4D Hypercube};
				\node[] at (3,14) {Single bar};
			\end{scope}
		
			\draw[-{Stealth[scale=2]}] (sr.east) -- (40, 3.5);
		\end{tikzpicture}
	\end{center}
	\vspace*{-0.2cm}
	\caption{The basic concept of our novel non-local hyperspectral reconstruction.}
	\label{fig:csr_concept}
\end{figure*}

Our novel algorithm Collaborative Spectral Reconstruction (CSR) introduces non-local block matching for the spectral reconstruction task in noisy environments and integrates a simultaneous and collaborative reconstruction procedure. Non-local block-matching and collaborative filtering became very popular with the introduction of the block-matching 3D filtering (BM3D) \cite{dabov-image-2007} grayscale denoiser. Since the state of the art only exploits the local neighborhood by block-processing procedures, non-local collaborative reconstruction has not found its way into spectral reconstruction until now.

Basically, the novel algorithm consists of three steps, which are shown in \fig\ref{fig:csr_concept}. First, similar multispectral cubes are found by a cube matching procedure. Afterwards, these multispectral cubes are stacked on top of each other, which results in a 4D hypercube. Second, the spectra are reconstructed along the new stacked dimension. These stacked multispectral pixels can be reconstructed collaboratively to reduce the influence of noise. Finally, since one pixel can be part of multiple 4D hypercubes, all reconstructed spectra belonging to the same pixel are averaged.

To reconstruct pixels non-locally a cube matching procedure is necessary. The distance used for the cube matching is a simple mean squared error between two multispectral cubes with size ${\BSRecon \times \BSRecon \times \NumMSChannels}$, and is calculated by
\begin{equation}
	\text{d}(\MSImageBlockVec^{\CoordOne}, \MSImageBlockVec^{\CoordTwo}) = ||\MSImageBlockVec^{\CoordOne} - \MSImageBlockVec^{\CoordTwo}||_2^2.
\end{equation}
These distances are used to build a 4D array $\MSCube^{\CoordOne}$ by stacking similar cubes to $\MSImageBlockVec^{\CoordOne}$ on top of each other. Consequently, $\MSCube^{\CoordOne}$ has the dimensions $\BSRecon\times\BSRecon\times\NumStackedBlocks\times\NumMSChannels$, where $\NumStackedBlocks$ is the number of stacked cubes. The hypercubes are sorted by ascending distance. Therefore, the cube $\MSImageBlockVec^{\CoordOne}$ itself with distance $d(\MSImageBlockVec^{\CoordOne}, \MSImageBlockVec^{\CoordOne}) = 0$ is always at the first index in dimension three of $\MSCube^{\CoordOne}$. The coordinates of the matched cubes are stored in $\BlockIndices^{\CoordOne}$.

Furthermore, a maximum number $\MaxBlocks$ or a threshold $\ThresholdBlocks$ on the distance limits the number of multispectral cubes in the 4D array. This threshold is calculated by $\ThresholdBlocks = \ThresholdBlocksConstant \cdot \overbar{\NoiseVar}$, where $\overbar{\NoiseVar}$ is the average noise variance over the multispectral images. This is done to control the number of similar cubes relative to the noise strength. Therefore, if there is no noise at all, and thus, the spectral reconstruction of a pixel does not need support from other pixels, stacking and filtering similar cubes would worsen the result. Consequently, the size of the stacking dimension is usually one in this case.

The next step is to reconstruct spectra out of multispectral pixels. An interesting reconstruction concept is the SSW \cite{murakami-color-2008}. The SSW exploits the correlation between pixels in the neighborhood. While the pixels are correlated in the neighborhood, the correlation might not always be high, especially in areas with edges. Here, the reconstruction acts in a non-local manner along the stacking dimension also using a Wiener filter. This exploits the fact that the same or very similar pixels are distributed across the whole image. As the noise acts independently on these pixels, this information can be used to lower the influence of noise on the reconstruction.

For the Wiener filter, the covariance matrix $\CorrelationMS$ of the multispectral channels is required as well as the cross-covariance matrix $\CorrelationHSMS$ between the multispectral channels and the hyperspectral channels. Consequently, the Wiener filter can be setup by the classic Wiener equation $\CorrelationHSMS \CorrelationMS^{-1}$. Using the extended version of the basic equation in \eqref{eq:basic_eq_extended} to calculate the covariance matrices, and exploiting the assumption that the AWGN has zero mean and is uncorrelated to $\HSChannelsBig$, leads to
\begin{equation}
	\CorrelationMS = \mathcal{E}\{ \left(\BigFilterMatrix \HSChannelsBig + \NoiseBig\right)\left(\BigFilterMatrix \HSChannelsBig + \NoiseBig\right)^{\text{T}} \} = \BigFilterMatrix\CorrelationSpatioSpectralCSR\BigFilterMatrix^{\text{T}} + \MSNoiseVarMatrixBig,
\end{equation}
where $\CorrelationSpatioSpectralCSR = \mathcal{E}\{\HSChannelsBig \HSChannelsBig^{\text{T}}\}$ is a spatio-spectral covariance matrix, which embeds the correlation in spectral dimension as well as in the stacked spatial dimension. However, the spatial correlation along the stacked dimension is an all-one matrix, since the pixels along this dimension are very similar due to the block-matching procedure. Thus, this matrix is built by $\CorrelationSpatioSpectralCSR = \mathbf{1} \otimes \CorrelationHS$. Matrix $\CorrelationHS$ is the spectral covariance matrix and can be derived by assuming the spectrum to be smooth. One way to express smoothness is by assuming the derivatives to be small. Since the spectrum to be estimated is sampled, differences instead of derivatives are considered. Thus, assuming the second order differences of the sampled spectrum to be small leads a smooth reconstructed spectrum.

The covariance matrix $\CorrelationHS$ can be derived from minimizing the second order differences. The optimization problem for the noiseless case with second order difference matrix $\DifferenceMatrix_2$ \cite{sippel-structure-preserving-2020}
\begin{equation}
	\begin{aligned}
		\ & \argmin_{\HSChannels} & & ||\DifferenceMatrix_2 \HSChannels||_2^2 \\
		\ & \text{s.t.} & & \MSChannels = \FilterMatrix \HSChannels
	\end{aligned}
\end{equation}
results in the filter
\begin{equation}
	\label{eq:sp_opt}
	\SmoothingMatrix^{-1}\FilterMatrix^{\text{T}}(\FilterMatrix\SmoothingMatrix^{-1}\FilterMatrix^{\text{T}})^{-1},
\end{equation}
where \mbox{$\SmoothingMatrix = \DifferenceMatrix_2^{\text{T}}\DifferenceMatrix_2 + \SmoothingMatrixInvFactor\Identity$} and $\SmoothingMatrixInvFactor$ is a small ($\alpha \ll 1$) constant since $\DifferenceMatrix_2^{\text{T}}\DifferenceMatrix_2$ is not invertible.  The second order difference matrix can be set up using the first order difference matrix 
\begin{equation}
	\begin{aligned}
		\DifferenceMatrix_1 &= \begin{pmatrix}
			1 & -1 & 0 & \cdots & 0 \\
			0 & 1 & -1 & \cdots & 0 \\
			\vdots & \ddots & \ddots & \ddots & \vdots  \\
			0 & \cdots & 0 & 1 & -1
		\end{pmatrix}
	\end{aligned}
\end{equation}
by calculating the difference of the difference which results in \mbox{$\DifferenceMatrix_2 = \DifferenceMatrix_1^{\NumHSChannels-2 \times \NumHSChannels-1} \DifferenceMatrix_1^{\NumHSChannels-1 \times \NumHSChannels}$}.

Applying the Wiener filter to a single pixel without noise results in
\begin{equation} 				
	\CorrelationHS\FilterMatrix^{\text{T}}(\FilterMatrix\CorrelationHS\FilterMatrix^{\text{T}})^{-1}.
\end{equation}
This is basically the same equation as in \eqref{eq:sp_opt}, where the covariance matrix $\CorrelationHS$ replaces the smoothing matrix $\SmoothingMatrix^{-1}$. Thus, the covariance matrix is set to $\CorrelationHS = \SmoothingMatrix^{-1}$.

The cross-covariance matrix between the multispectral channels and the hyperspectral channels is calculated by
\begin{equation}
	\CorrelationHSMS = \mathcal{E}\{\HSChannelsBig \left(\BigFilterMatrix \HSChannelsBig + \NoiseBig\right)^{\text{T}} \} = \CorrelationSpatioSpectralCSR\BigFilterMatrix^{\text{T}},
\end{equation}
where again the assumptions about AWGN are exploited. Using both covariance matrices $\CorrelationHSMS$ and $\CorrelationHS$ leads to the reconstruction
\begin{equation}
	\HSCube^{\CoordOne}(\CoordTwo) = \CorrelationSpatioSpectralCSR\BigFilterMatrix^{\text{T}}(\BigFilterMatrix\CorrelationSpatioSpectralCSR\BigFilterMatrix^{\text{T}} + \MSNoiseVarMatrixBig)^{-1}\text{vec}(\MSCube^{\CoordOne}(\CoordTwo)),
\end{equation}
where $\CoordTwo$ is a two dimensional coordinate for the first two dimensions and $\text{vec}(\cdot)$ vectorizes the 2D array.

In the end, since the same pixel is often part of multiple $\HSCube(\CoordOne)$ as depicted in \fig\ref{fig:csr_concept}, all spectra corresponding to the same pixel need to be mixed. For this, a simple average over the corresponding pixels is calculated. To simplify the formulation, each hyperspectral slice is extended outside its block by zero according to its position in the image, which is stored in $\BlockIndices^{\CoordOne}$, such that the first two dimensions match the dimension of the hyperspectral image. These estimates of the hyperspectral images are denoted as $\HSCubeExtended^{\CoordOne}$. Therefore, the averaging can be described by
\begin{equation}
	\HSImageEst^{\text{CSR}}(\CoordOne) = \frac{\sum_{\CoordTwo \in \ImageCoords}\sum_{i=1}^{|\BlockIndices^{\CoordTwo}|}\HSCubeExtended^{\CoordTwo}(\CoordOne, i)}{\sum_{\CoordTwo \in \ImageCoords}\sum_{i=1}^{|\BlockIndices^{\CoordTwo}|}\Indicator(\HSCubeExtended^{\CoordTwo}(\CoordOne, i))},
\end{equation}
where $\ImageCoords$ contains all coordinates of pixels that own a hypercube $\MSCube^{\CoordOne}$ and $\Indicator(\HSCubeExtended^{\CoordTwo}(\CoordOne, i))$ indicates when $\CoordOne$ is inside the reconstructed block area of the $i$-th slice of $\HSCubeExtended^{\CoordTwo}$. Thus, if $\CoordOne$ is in the extended area with value zero, the indicator function is zero. Consequently, the numerator basically sums up all extended slices, while the denominator counts the number of reconstructed spectra for every pixel.

\section{Evaluation}
\label{sec:evaluation}

\begin{figure}[t]
	\begin{center}
		\input{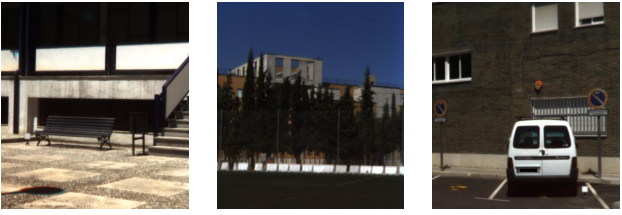}
	\end{center}
	\vspace*{-0.2cm}
	\caption{RGB images simulated from the hyperspectral database \cite{eckhard-outdoor-2015}.}
	\label{fig:rgb_images}
\end{figure}

\begin{figure}[t]
	\begin{center}
		\input{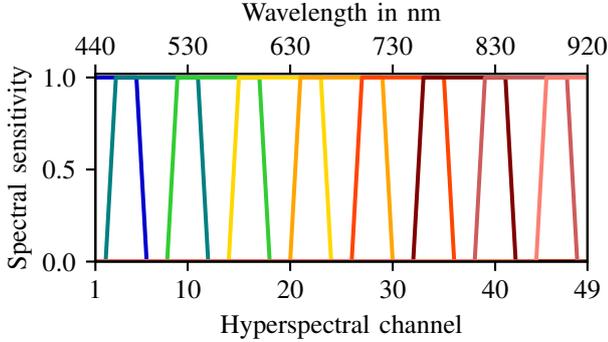}
	\end{center}
	\vspace*{-0.2cm}
	\caption{Nine synthetic filters to simulate the integration process of a multispectral camera.}
	\label{fig:filters}
\end{figure}

\begin{figure*}[t]
	\begin{center}
		\input{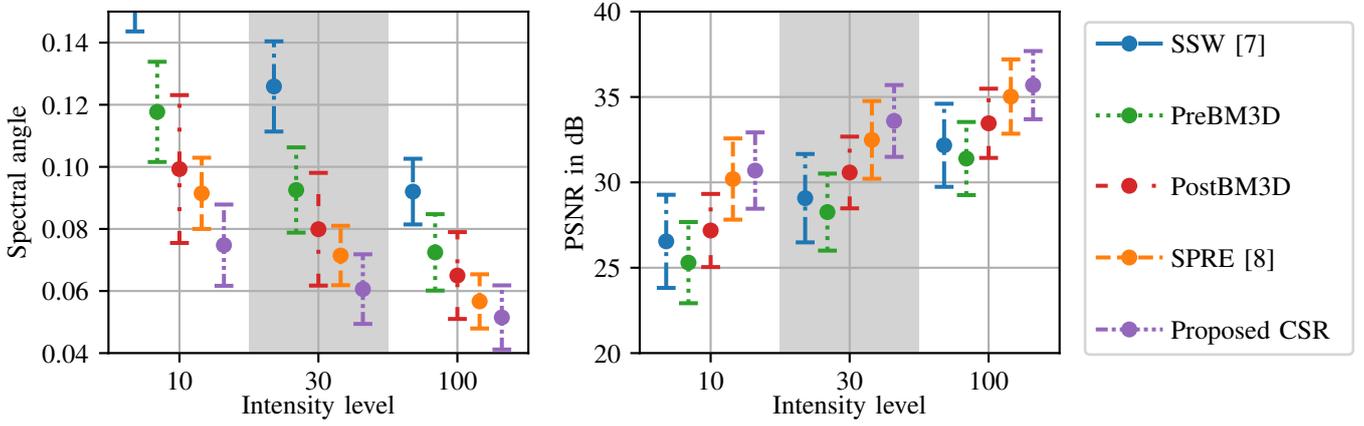}
	\end{center}
	\vspace*{-0.2cm}
	\caption{The spectral angle $\theta$ and the PSNR averaged over all hyperspectral images in the database for three different scenarios with intensity level 10, 30 and 100, respectively. The lower the spectral angle, the better the reconstruction. The big circles indicate the mean, while the bars show plus/minus the standard deviation.}
	\label{fig:reco_results}
\end{figure*}

For the evaluation, we use a database containing 14 urban hyperspectral images \cite{eckhard-outdoor-2015}. Since some of the spectral bands are corrupted by strong noise, the wavelengths from 440 nm to 920 nm are used. Sampling this area in 10 nm steps results in 49 hyperspectral channels. Examples of simulated RGB images from this database are shown in \fig\ref{fig:rgb_images}.

To generate multispectral data out of these hyperspectral images, synthetic filters are necessary. These filters are shown in \fig\ref{fig:filters}. The nine filters are of a uniform flat-top shape and have a considerable amount of overlap.

Since the most important noise source is shot noise, Poisson distributed noise is added to the multispectral images. Specifically, the noise distribution
\begin{equation}
	\label{eq:poisson_noise}
	\MSImage_i(\CoordOne) \sim \frac{\mathcal{P}\left(\IntensityLevel \ \MSImageNoiseless_i(\CoordOne) \right)}{\IntensityLevel}
\end{equation}
is used, where $\MSImageNoiseless_i(\CoordOne)$ is the noiseless multispectral image and $\IntensityLevel$ indicates the intensity level. A low intensity level corresponds to a low amount of photons, and thus, to a high amount of noise. While simulating the noise, the values of the multispectral image are normalized to its maximum. Thus, all values are between zero and one. It has to be noted that the proposed method was modeled by assuming Gaussian noise, as this leads to a closed-form solution, while the evaluation is based on Poisson noise, which is the appropriate model for shot noise. Thus, the evaluation also picks up how well the different methods generalize to this different type of noise.

The reconstructed spectra are evaluated using the spectral angle $\theta$ as metric. The spectral angle $\theta$ is defined by \cite{kruse-spectral-1993}
\begin{equation}
	\theta(\HSChannels, \HSChannelsEst) = \arccos{\left(\frac{\HSChannels^{\text{T}}}{||\HSChannels||_2}\frac{\HSChannelsEst}{||\HSChannelsEst||_2}\right)}.
\end{equation}
and, a lower spectral angle denotes a better reconstruction. The worst result a reconstructed spectrum can have is $\pi$. $\theta(\HSChannels, \HSChannelsEst)$ is averaged over the whole image. Original pixels with an all-zero spectrum do not contribute to $\theta(\HSChannels, \HSChannelsEst)$, while all-zero reconstructed spectra are contributing to $\theta(\HSChannels, \HSChannelsEst)$ with the value $\pi$.

Furthermore, the peak signal-to-noise ratio (PSNR) is used for the evaluation
\begin{equation}
	\text{PSNR}(\HSImage, \HSImageEst) = -10 \cdot \log_{10}\left(\frac{1}{\NumHSChannels  \ImageWidth \ImageHeight}||\HSImage - \HSImageEst||_2^2\right),
\end{equation}
where the maximum value of the hyperspectral images is one, and $\ImageWidth$ and $\ImageHeight$ is the image width and height, respectively.


Our novel method CSR is compared to the SSW \cite{murakami-color-2008} and the state-of-the-art algorithm structure-preserving reflectance estimation (SPRE) \cite{sippel-structure-preserving-2020} as well as to the application of the BM3D \cite{dabov-image-2007}. Specifically, the multispectral image are denoised using BM3D first and the reconstruction is done using a Wiener filter \cite{pratt-spectral-1976} afterwards (PreBM3D), and vice versa, namely, first the reconstruction with a Wiener filter and afterwards the application of the BM3D (PostBM3D). All parameters of the competitors are set to the ones of the corresponding papers. The block size $\BSCSR$ for our CSR is 8. The search window in which similar multispectral cubes are found has a size of $33\times33$. To enhance performance not every pixel owns a stacked 4D cube. The step size to the next pixel in $\ImageCoords$ is 3 in horizontal and vertical direction. These values are derived from the values of the BM3D. $\ThresholdBlocksConstant$ has the value 6, which was determined using the independent CAVE database \cite{yasuma-generalized-2010}, and $\MaxBlocks$ is set to 25. The amount of pixels that are simultaneously reconstructed matches with the SSW and SPRE, where the block size is 5. Thus, using this block size and $\MaxBlocks$ results in a balanced competition. The maximum number of stacked blocks for the BM3D is set to 32, since this needs to be an exponential number with base 2. Since PreBM3D and PostBM3D should not be handicapped, this parameter is set to the closest possible number that is greater than 25. The multispectral images are padded beforehand according to the requirements of the reconstruction method. It is noteworthy that the results of CSR may get even better with optimized parameters. However, the parameters are chosen such that a fair comparison is possible.

The results are shown in \fig\ref{fig:reco_results}. Our novel CSR outperforms all algorithms for all three noise scenarios. The PSNR is improved by up to 1.11dB and the spectral angle up to 18\%. However, the standard deviation is slightly higher in comparison to SPRE.
\begin{figure*}[tb]
	\begin{center}
		\input{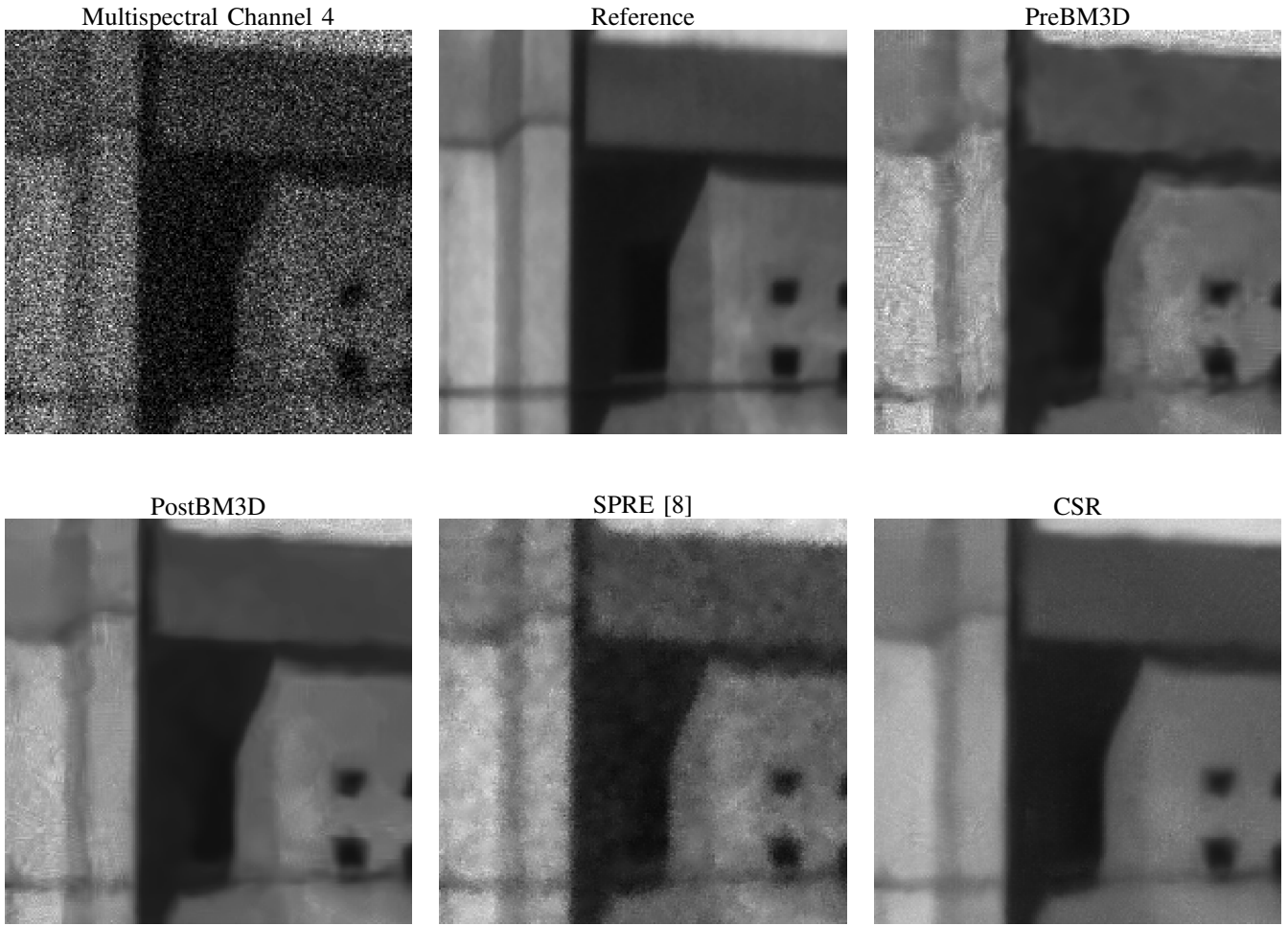}
	\end{center}
	\vspace*{-0.2cm}
	\caption{A $200 \times 200$ area of channel 20 of a reconstructed image for intensity level 10. In the top left corner channel four of the noisy multispectral image is shown. The SSW is omitted due to its bad performance.}
	\label{fig:reco_images}
\end{figure*}

\begin{figure*}[h!]
	\begin{center}
		\input{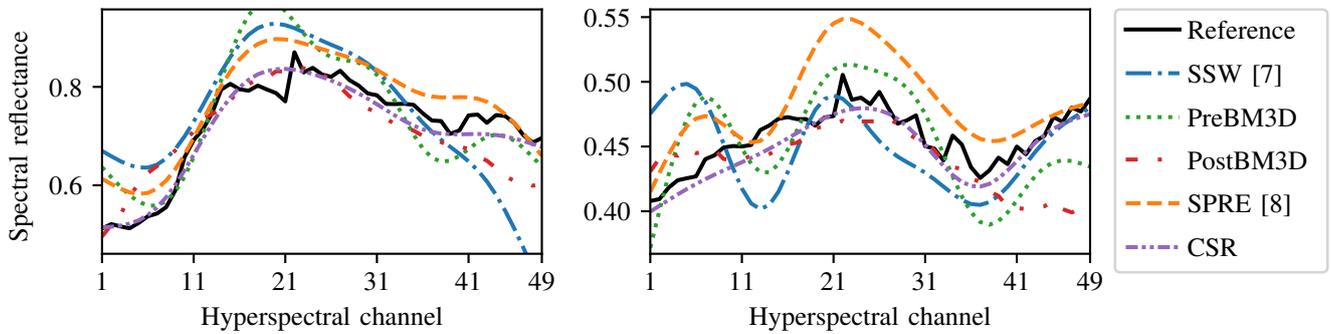}
	\end{center}
	\vspace*{-0.4cm}
	\caption{Reconstructed spectra of two different pixels with intensity level 10.}
	\label{fig:spectra}
	\vspace*{-0.4cm}
\end{figure*}

\fig\ref{fig:reco_images} shows a reconstructed hyperspectral channel for nearly all algorithms as well as channel four of the noisy multispectral image to get an impression of the amount of Poisson noise added to the multispectral images. Furthermore, the ground-truth hyperspectral channel is also included. The visual result of our novel method is more appealing than the images of the other three algorithms, since it preserves sharp structures while being much less noisy in homogeneous regions. This is particularly visible in the left part of the shown image. Interestingly, while PostBM3D calculates visually more appealing images, SPRE reconstructs more accurate spectra. Our novel CSR returns both, visually appealing images and accurate spectra. \fig\ref{fig:spectra} shows reconstructed spectra of a single pixel. While the SPRE also delivers usable spectra, our novel algorithm reconstructs spectra even better.

\section{Conclusion}
\label{sec:conclusion}

A novel reconstruction method called Collaborative Spectral Reconstruction (CSR) was introduced. This new algorithm uses a global block-matching procedure on the noisy multispectral images. Subsequently, the spectra are collaboratively reconstructed along the matched dimension using a spatio-spectral Wiener filter which assumes the second order difference of the spectrum to be small. In comparison to the original spatio-spectral Wiener filter, our novel approach reconstructs multiple multispectral pixels in non-local fashion at once, which exploits the fact that the same or very similar pixels are distributed all over the image while the noise acts independently on these pixels. The evaluation showed that our novel algorithm outperforms the state of the art in terms of the spectral angle, which is improved up to 18\% in noisy scenarios, and PSNR, which is raised by 1.11dB, as well as from a visual point of view. Further work will address the question which filter selection performs best considering reconstruction quality and noise.

\bibliographystyle{ieeetr}
\bibliography{refs}
\vspace{12pt}

\end{document}